\documentclass[aps,prl,twocolumn,showpacs,superscriptaddress]{revtex4-1}  % for review and submission
\usepackage{graphicx}
\usepackage{color}
\usepackage{amsmath}
\usepackage{hyperref}
\usepackage{ulem}

\usepackage{float}
\usepackage[utf8]{inputenc}
\usepackage{placeins}

\begin{document}

\title{Fine-tuning the DNA conductance by intercalation of drug molecules}

\author{Abhishek Aggarwal}
\affiliation{Center for Condensed Matter Theory, Department of Physics, Indian Institute of Science, Bangalore-560012, India}

%\author{Anil Kumar Sahoo}
%\affiliation{Center for Condensed Matter Theory, Department of Physics, Indian Institute of Science, Bangalore-560012, India}

\author{Anil Kumar Sahoo}
\thanks{Present address: Biomaterials Department, Max Planck Institute of Colloids and Interfaces, 14476 Potsdam, Germany \newline
Fachbereich Physik, Freie Universit{\"a}t Berlin, 14195 Berlin, Germany}
\affiliation{Center for Condensed Matter Theory, Department of Physics, Indian Institute of Science, Bangalore-560012, India}

\author{Saientan Bag}
\affiliation{Center for Condensed Matter Theory, Department of Physics, Indian Institute of Science, Bangalore-560012, India}

\author{Veerabhadrarao Kaliginedi}
\affiliation{Department of Inorganic and Physical Chemistry, Indian Institute of Science, Bangalore 560012}

\author{Manish Jain}
\affiliation{Center for Condensed Matter Theory, Department of Physics, Indian Institute of Science, Bangalore-560012, India}

\author{Prabal K. Maiti}
\email{Corresponding author: maiti@iisc.ac.in}
\affiliation{Center for Condensed Matter Theory, Department of Physics, Indian Institute of Science, Bangalore-560012, India}      
\date{\today}

\begin{abstract}
In this letter, we study the structure-transport property relationships of small ligand intercalated DNA molecules using a multiscale modelling approach where extensive ab-initio calculations are performed on numerous MD-simulated configurations of dsDNA and dsDNA intercalated with two different intercalators, ethidium and daunomycin.
DNA conductance is found to increase by one order of magnitude upon drug intercalation due to the local unwinding of the DNA base pairs adjacent to the intercalated sites which leads to modifications of the density-of-states in the near-Fermi energy region of the ligand--DNA complex. 
Our study suggests that the intercalators can be used to enhance/tune the DNA conductance which opens new possibilities for their potential applications in nanoelectronics.
\end{abstract}

\maketitle

DNA intercalators have been a subject of intense scientific research because of their various uses, such as anticancer and antitumor drugs\cite{hurley2002dna} and fluorescent tags in imaging\cite{backer2019single}. The molecular mechanism of the ligand intercalation process, especially the kinetics and thermodynamics of ligand intercalation have been well studied\cite{almaqwashi2016mechanisms, wilhelm2012multistep, mukherjee2008molecular, chaires1997energetics, aggarwal2020we}. Recently, many experimental studies have focused on understanding how intercalators modify the mechanical properties of double-stranded DNA (dsDNA)\cite{almaqwashi2016mechanisms, stassi2019nanomechanical, schakenraad2017hyperstretching, lipfert2010torsional, burmistrova2016force, camunas2015single, gunther2010mechanical, krueger2016modeling, sahoo2019understanding}, inferring how intercalators could affect many active biological processes, such as DNA repair, replication, and transcription. However, despite immense biological and technological implications, only a few recent experimental studies have investigated the effect of ligand intercalation on DNA conductance\cite{harashima2017single,guo2016molecular, wang2015revealing}.

DNA has emerged as an integral part of molecular electronics over the past decade\cite{de2019nanomaterial, genereux2010mechanisms}. Several theoretical and experimental works have been done to explore the charge transport properties of DNA\cite{bag2016dramatic, aggarwal2018remarkable,xiang2017gate, li2016thermoelectric, bruot2015tuning, artes2015conformational, qi2013unified, song2008anomalous, zhuravel2020backbone, wolter2017microsecond, woiczikowski2009combined, troisi2002hole, artes2014nanoscale, prytkova2007photoselected, wolter2013charge, aggarwal2020multiscale, bag2020machine}. \textcolor{black}{Apart from molecular electronics, DNA charge transport has applications in spin specific electron conductor\cite{gohler2011spin}, and detection of genetic materials from an ensemble\cite{li2018detection}. DNA charge transport also has relevance} in various biological processes, such as redox switching of [4Fe4S] clusters found in all DNA processing enzymes, which in turn affects DNA repair and replication processes\cite{o20174fe4s, bartels2017electrochemistry}. DNA structure is highly distorted in the process of ligand intercalation, in which the planar aromatic rings of a ligand intercalate between two successive DNA base pairs\cite{galindo2015intercalation, aggarwal2020we}, significantly affecting the charge transport in DNA. 

Recently, using the STM-BJ technique, Harashima et al.\cite{harashima2017single}  have studied the effect of intercalation and groove binding on the conductance of a 8 base pairs (bp) long dsDNA and found that the DNA conductance increases by almost four times when an ethidium is intercalated into the DNA, whereas the conductance remains unchanged for groove binding\cite{harashima2017single}. Guo et al.\cite{guo2016molecular} have shown the rectification behaviour of DNA upon the intercalation of coralyne molecules. Wang et al.\cite{wang2015revealing} have studied the change in conductance of dsDNA upon the intercalation of SYBR green and ethidium bromide (EB) and found that the DNA device conductance decreases upon treatment with EB. Liu et al.\cite{liu2011direct} measured the conductance of metallo-DNA complexes and  efficiently switched on-and-off their electrical properties. However, the physics behind the structural changes due to the ligand--DNA intercalations and their effect on the charge transport mechanism remains unknown, and a coherent understanding of structure-transport relationships for the ligand--DNA intercalations is yet to be established. This calls for theoretical investigations.

\begin{figure}[htbp]
    \centering
        \includegraphics[width=\columnwidth,height=1.15\columnwidth]{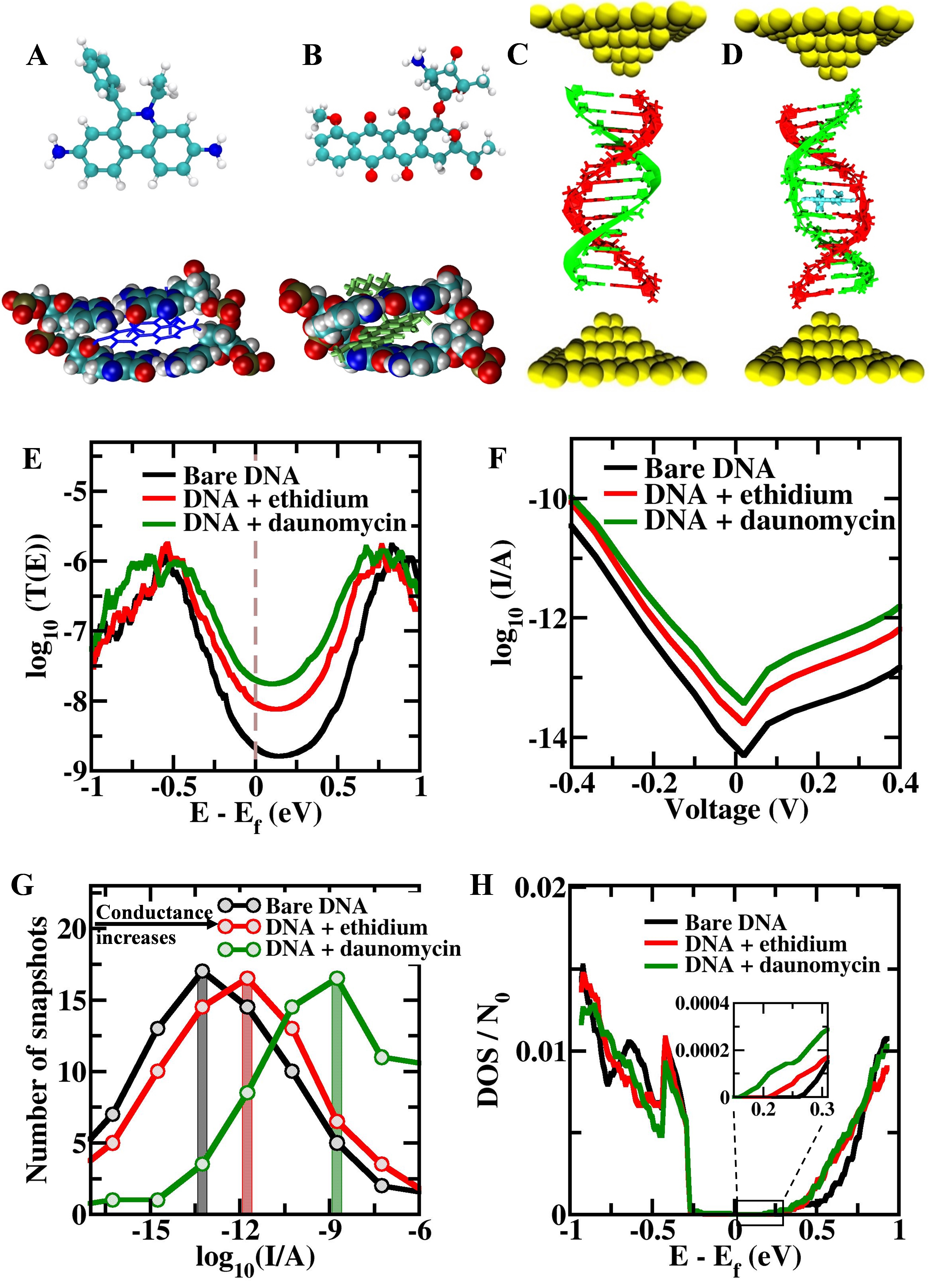}
	\caption{Atomic Structure and intercalated arrangement of A) ethidium (blue colored) and B) daunomycin (green colored) between two base pairs (shown in VDW representation). Schematic representations showing the charge transport set-up and structure of C) bare dsDNA: two strands shown in green and red color respectively and D) ethidium intercalated dsDNA: the ethidium (blue colored) intercalated between two base pairs of dsDNA. The \textcolor{black}{virtual} gold electrodes are shown as yellow spheres, while water molecules and ions are not shown here for clarity. E) Transmission probability curve for the DNA and drug-DNA complexes (averaged over 75 morphologies) in the region close to the Fermi-energy for 8 bp dsDNA sequence with and without intercalators. F) V-I characteristics curves of 8 bp dsDNA with and without intercalators. G) Distribution of the log of current at an applied potential of 100 mV for bare dsDNA and intercalated dsDNA. The intercalated dsDNA has higher number of snapshots for larger current value than the bare dsDNA. H) Density-of-states for the 8 bp dsDNA in the presence and absence of an intercalator \textcolor{black}{computed using the energy states of all the 75 morphologies studied for each case}. The inset shows the zoomed view DOS in the positive side of Fermi-energy region. }
    \label{fig:1}
\end{figure}

The typical theoretical DNA charge-transport works consist of ab-initio calculations performed on a single optimized structure of the molecule\cite{artes2015conformational, mishchenko2010influence, li2018detection, valianti2019charge, seth2017conductance}, but that does not capture the real essence of experimental studies such as break-junction experiments\cite{rascon2015binding}. A lot of factors arising due to the fluctuations in the geometry of the molecule such as, attachment geometry of molecule to the electrodes\cite{pal2018electronic}, intramolecular tilt/twist angles and conformation of the molecular bridge, have a huge impact on the conductivity of the single-molecule junctions\cite{nichols2010experimental, wen2013electrical, su2016chemical}. To consider these fluctuations in the conductivity of the molecule, a methodology is required which captures the randomness of the system at a molecular level. The previous works in which multiple morphologies of the DNA systems are used to study the role of fluctuations in the charge transport efficiencies either focus only on particular base pairs of DNA leaving the terminal base pairs from the calculations\cite{troisi2002hole, troisi2003rate, berlin2008charge, gutierrez2009charge} or use tight-binding approximations to compute the Hamiltonian of the system\cite{berlin2008charge, gutierrez2009charge, grozema2008effect, gutierrez2010structural}. However, ab-initio calculations for the full DNA systems are not performed because of being computationally expensive\cite{gutierrez2010structural, kubavr2013efficient}. Here, we have used a multiscale modelling approach which couples classical all-atom molecular dynamics (MD) simulations, extensive quantum mechanical calculations, and non-equilibrium Green’s function (NEGF) methods on full DNA/ drug-DNA complex systems. By employing this methodology to at least 75 MD-simulation-sampled structures, we explain the physics behind the difference in the charge transport properties of a bare dsDNA and the intercalated ligand--DNA complexes for two different intercalators namely, ethidium and daunomycin (Fig. \ref{fig:1}A-B).  

We extend the DNA strands of the crystal structures of ethidium and/or daunomycin intercalated dsDNA complexes \cite{jain1984visualization, wang1987interactions} to build 12 and 8 bp long dsDNAs of sequences (GCGCA$\bf{CG}$TGCGC)$_{2}$ and (GCA$\bf{CG}$TGC)$_{2}$, keeping an intercalator (daunomycin or ethidium) between the middle two base pairs (bold characters in the sequences) as shown in Fig. \ref{fig:1}. For the above 8 bp long dsDNA, we increase the concentration of ethidium as well as  intercalate ethidium at asymmetric positions of the DNA sequence,  to build (G$\bf{CA}$CG$\bf{TG}$C)$_{2}$, (G$\bf{CA}$CGTGC)$_{2}$ and (GCACG$\bf{TG}$C)$_{2}$. Additionally, for a direct comparison with the experiment\cite{harashima2017single}, we build an 8 bp long dsDNA of sequence (GCT$\bf{TG}$TTG)$_{2}$ in the presence of an intercalated ethidium molecule. We follow, here, the same all-atom MD simulation protocol as described in our earlier publication\cite{sahoo2019understanding} and Supplementary Information (SI) to simulate dsDNA and drug-dsDNA complexes for 200 ns. The conductance of the dsDNA molecules is computed using Landauer formalism \textcolor{black}{where the electrodes are modelled virtually using a coupling parameter. In section S4 of SI, we show that the choice of this coupling parameter affects the results only quantitatively, qualitatively the results remain same. Readers are referred to section S3 of SI for a discussion on the validity of Landauer formalism on the systems studied in this work}. For each complex, the Landauer formalism is applied to 75 MD-simulation-sampled structures to get the average I-V characteristics.

\begin{figure}[htbp]
    \centering
	\includegraphics[width=\columnwidth]{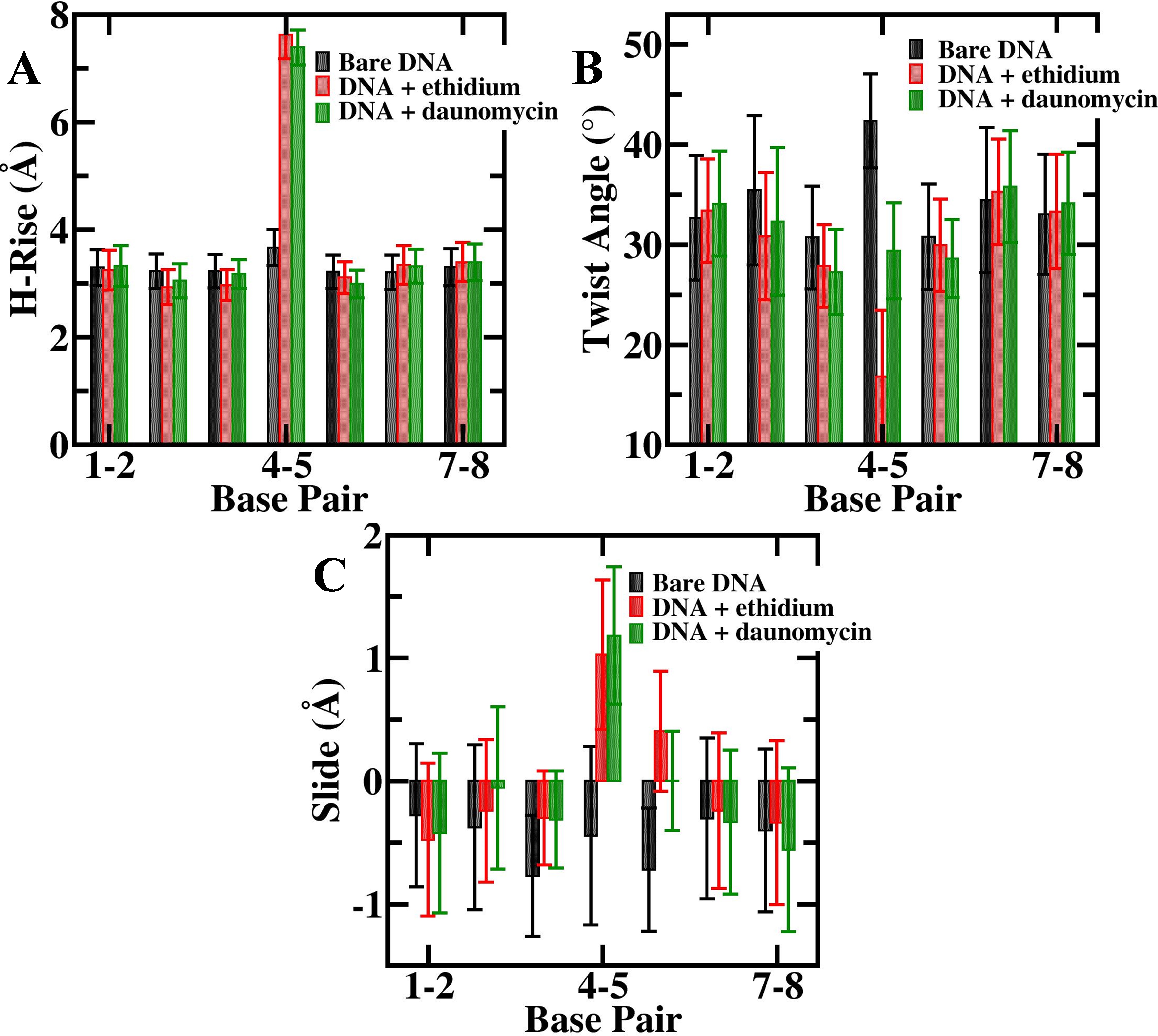}
    \caption{Structural parameters of a bare 8 bp dsDNA and the same dsDNA with an ethidium or a daunomycin intercalated between the middle two base pairs. The height of the bars represents the average value of the parameters while the error bars denote their standard deviation calculated using the last 50 ns of the 200 ns long trajectory.}
    \label{fig:2}
\end{figure}

To investigate how intercalators modify the dsDNA structure, we calculate different inter-base pair helical parameters of the bare dsDNA as well as intercalated dsDNA complexes. The results for the 8 bp and 12 bp long dsDNA are shown in Fig. \ref{fig:2} and Fig. S4,S5 in the SI, respectively. For each of the intercalated ligand--DNA complexes, the rise of the base pair step at the intercalation site is almost double than that of the bare dsDNA. The rise is similar for rest of the base pairs irrespective of the presence of an intercalator. The slide also varies in the intercalated region. There is a notable difference in the twist angle parameter for the bare dsDNA and intercalated dsDNA. The magnitude of twist angle for the intercalated region is significantly less than that of the bare dsDNA. This signifies the local unwinding of the base pairs in the intercalated region upon intercalation. This leads to significant change in the relative orientation of the base pairs in the intercalated region and the base pairs become aligned to each other. Thus, the base pairs adjacent to the intercalators have high rise but at the same time a lower twist angle than the corresponding bare dsDNA base pairs.

To understand the effects of these structural changes of dsDNA upon intercalation on the charge transport properties of dsDNA, we computed the transmission probabilities for 8 long dsDNA. Fig. \ref{fig:1}E shows the comparison of transmission probabilities for the 8 bp dsDNA intercalated with ethidium or daunomycin and the bare 8 bp dsDNA for a range of energies near the Fermi energy of the molecule. Here, fluctuations in the transmission probabilities, under equilibrium condition, are averaged on the log-scale, as the distribution of the tunneling conductance is expected to be log-normal\cite{venkatramani2014breaking}.  Clearly, the intercalated dsDNA has higher transmission relative to the bare dsDNA for both the intercalating ligands, resulting in  their higher conductance  than the bare dsDNA (Fig. \ref{fig:1}F). The daunomycin--dsDNA complex has almost one order of magnitude higher conductance than the bare dsDNA. Fig. \ref{fig:1}G compares the distribution of the number of snapshots of bare dsDNA and intercalated dsDNA for the current at an applied potential bias of 100 mV. Clearly, both ethidium as well as daunomycin intercalated dsDNA complexes have higher number of more conductive morphologies or active conformations\cite{woiczikowski2009combined} than the bare dsDNA. \textcolor{black}{These conformations show current of the order of microamperes. Hence, an arithmetic mean here will lead to an average current of the order of microamperes which is in line with experimental observations\cite{harashima2017single}. However, to provide a clear picture which represents the contribution of all the snapshots judiciously, we present log-average of the currents in all the I-V characteristic graphs}. We also computed the coherent charge transport properties for 12 bp DNA sequence. Like the 8 bp DNA, we find that the conductance of 12 bp DNA also increases upon intercalation of an ethidium or a daunomycin (Fig. S6 in the SI). \textcolor{black}{We also find that the dsDNA backbones play a crucial role in determining the electrical properties of dsDNA as discussed in section S8 of SI.}

The physics behind the increase in the transmission probability upon drug intercalation can be understood from the electronic density-of-states (DOS) in the region close to Fermi energy for the 8 bp dsDNA as shown in Fig. \ref{fig:1}H. The DOS curve shows that upon intercalation, the HOMO-LUMO gap gets smaller in magnitude compared to that of the bare dsDNA. This means that there are more energy states available for the charge conduction in the region near Fermi energy in the drug-intercalated dsDNA than in bare dsDNA. Quantitatively, the average HOMO-LUMO gap for bare dsDNA is 1.07 $\pm$ 0.14 eV while for ethidium-dsDNA and daunomycin--dsDNA complexes, it is 1.02 $\pm$ 0.15 eV and 0.99 $\pm$ 0.14 eV, respectively. This is evident from the inset of fig. \ref{fig:1}H. Notably, there are more DOS on the positive side of Fermi energy for the drug intercalated dsDNA relative to bare dsDNA. This feature is reflected in the transmission probabilities curve as well (Fig. \ref{fig:1}E), i.e. the transmission increases for lower energies in the case of intercalated dsDNA relative to that of the bare. 

\begin{figure}[htbp]
    \centering
        \includegraphics[width=0.95\columnwidth, keepaspectratio]{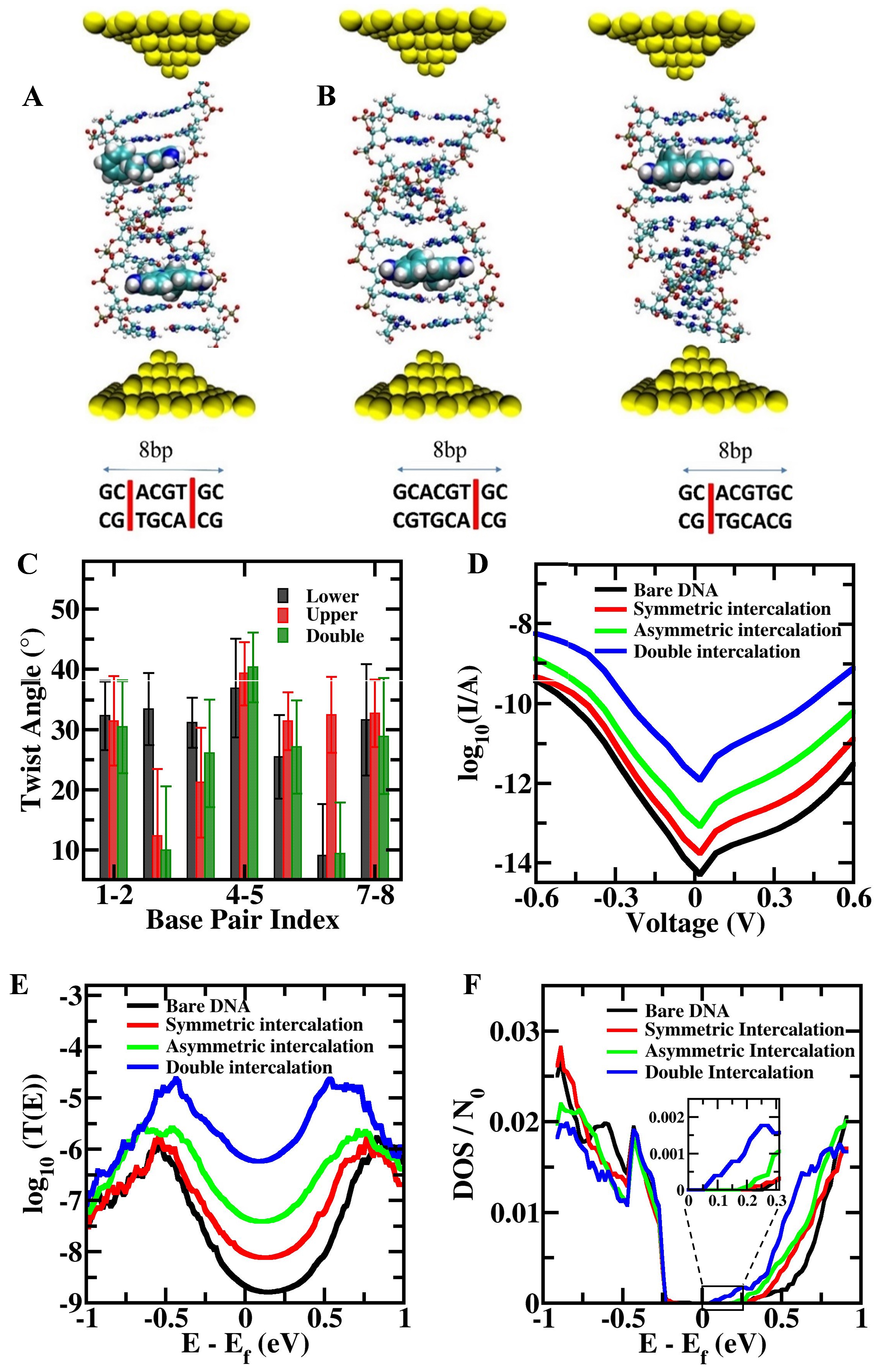}
	\caption{Schematic diagram of 8 bp dsDNA intercalated with ethidium intercalators at A) two different sites, B) at asymmetric positions of the dsDNA, i.e. at 2nd position and 6th position from top. C) The twist angle profile of asymmetrically intercalated and doubly intercalated dsDNA molecules. D) Comparison of transmission probabilities of the bare dsDNA with intercalated dsDNA. E) V-I characteristics curve for dsDNA intercalated with different number of ethidium at different intercalation sites. F) The DOS profile for the same systems as in A) and B).}
    \label{fig:3}
\end{figure}

To further check the robustness of our result, we studied charge transport through dsDNA by increasing the concentration of ethidium (see Fig. \ref{fig:3}A) as well as by intercalating ethidium at asymmetric positions of the DNA sequence (see Fig. \ref{fig:3}B). The twist angle shows a clear dip at the intercalated sites just like in the symmetric intercalation case as shown in Fig. \ref{fig:3}C. Fig. \ref{fig:3}D shows the comparison of the transmission probabilities curves for the asymmetric and symmetric intercalations. In the region close to Fermi energy, the transmission probabilities for the doubly intercalated dsDNA is one order-of-magnitude higher than that of the singly intercalated dsDNA. This highlights the fact that intercalating a dsDNA changes the dsDNA structure in such a way that it becomes more conductive electrically. Like the symmetrically intercalated dsDNA, the transmission probabilities for the asymmetrically intercalated dsDNA is higher relative to the bare dsDNA but lower than the doubly intercalated dsDNA. This signifies the generality of the finding that dsDNA conductance increases upon intercalation, independent of the drug intercalation site. Fig. \ref{fig:3}E shows the comparison of conductance of the bare dsDNA and the dsDNA--ethidium complexes. Increasing the concentration of intercalators in a dsDNA increases its conductance, as the singly intercalated dsDNA shows the intermediate conductance to the bare dsDNA and the double intercalated dsDNA. Fig. \ref{fig:3}F shows the comparison of the DOS of the bare dsDNA and the dsDNA--ethidium complexes. Clearly, the DOS in the region close to the Fermi energy for the singly (both symmetric and asymmetric) intercalated dsDNA is higher than that of the bare dsDNA but is lower than the doubly intercalated dsDNA. This trend of the DOS for the different molecules correlates well with the respective trend of the dsDNA conductance and transmission properties. These results signify the robustness of the increase in the dsDNA conductance upon drug intercalation.

\begin{figure}[htbp]
    \centering
        \includegraphics[width=\columnwidth,height=1.15\columnwidth]{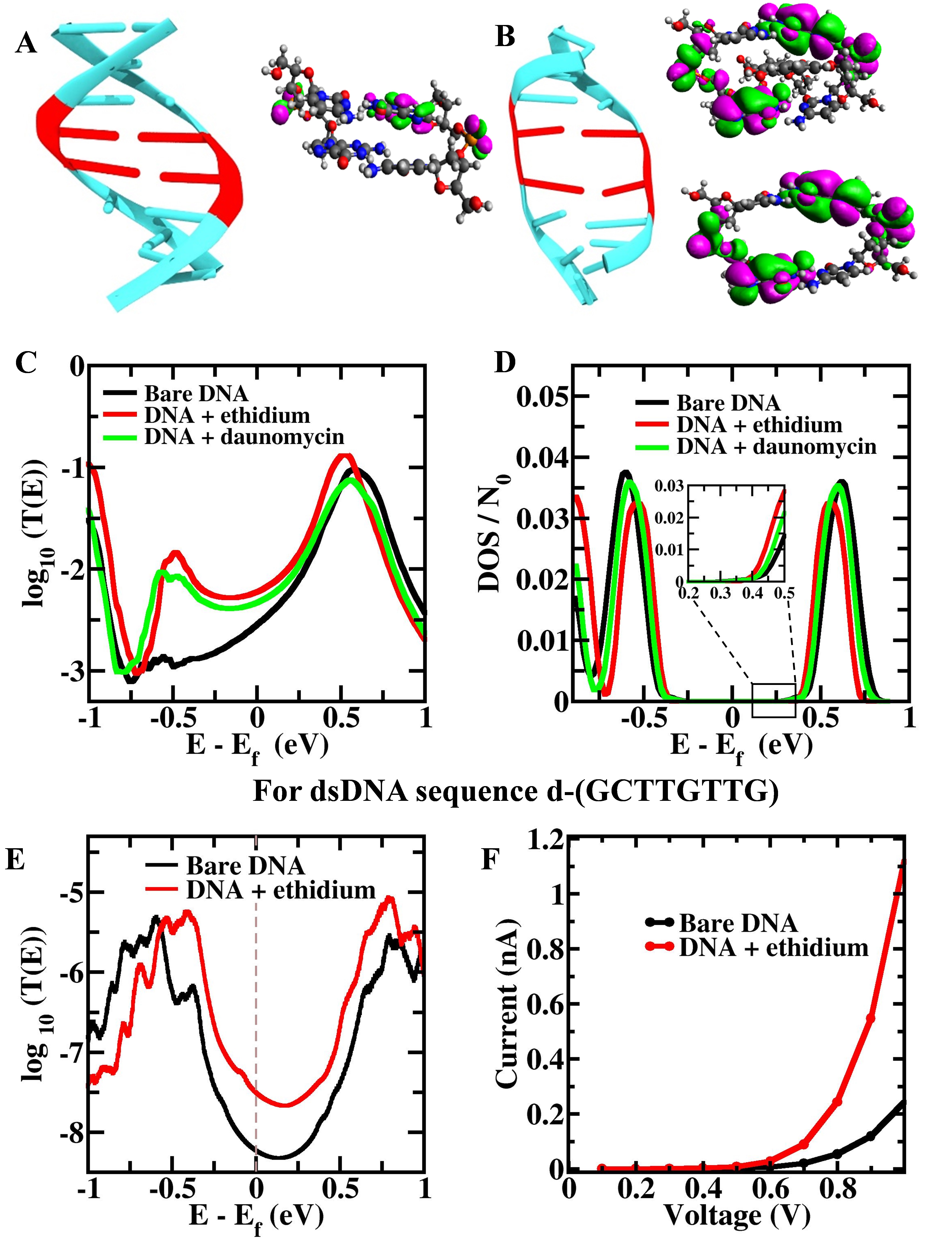}
	\caption{Schematic diagram highlighting the base pairs of A) bare dsDNA B) ethidium/daunomycin intercalated dsDNA. \textcolor{black}{The top part of B) represents the HOMO isosurface for the intercalated dsDNA part with and without intercalator included in the calculation. The isosurfaces are similar and the intercalator does not change the HOMO distribution significantly.} C) A comparison of the average transmission probabilities of the bare dsDNA and the intercalated dsDNA complexes in the region close to the Fermi energy. D) The Density of States (DOS) for only the intercalated region. E) Transmission probability curves in the region near the Fermi energy and F) V-I characteristics curves for the experimentally studied 8 bp dsDNA with sequence d-(GCTTGTTG) in the presence and absence of ethidium.}
	\label{fig:4}
\end{figure}

To better understand the reason behind the enhancement of DNA conductance upon drug intercalation, we have also calculated the transmission probability as well as electronic density of states for the intercalated region of the dsDNA. This part should capture the important physics behind the transmission through the whole dsDNA. For a one-to-one comparison with bare dsDNA, we have just considered the two adjacent base pairs to the intercalator without the intercalators and the corresponding base pairs of the bare dsDNA. Figs. \ref{fig:4}A and \ref{fig:4}B highlight the two base pairs considered for the transmission calculations. Clearly, the distance between the two base pairs (7.6 $\pm$ 0.4 {\AA}) in intercalated dsDNA is higher than that of the bare dsDNA (3.6 $\pm$ 0.3 {\AA}), but at the same time, the intercalated base pairs are more aligned geometrically than bare dsDNA base pairs. Note the significant decrease in the twist angle upon drug intercalation as shown in Fig. \ref{fig:2}E. The twist angle has also been found to affect the tunneling conductance of other single-molecule junctions\cite{venkataraman2006dependence, mishchenko2010influence}. Figures \ref{fig:4} A-B show the HOMO distribution on the intercalated base pairs. Clearly, for the intercalated base pairs, the HOMO distribution does not differ much regardless of whether the intercalator is considered into the calculations. This shows that the intercalator energy states do not lie near the HOMO level of the intercalated dsDNA and hence just considering the two adjacent base pairs to the intercalator without the intercalators should provide the fundamental understanding of the process. Fig. \ref{fig:4}C shows the transmission probability curves for these systems, where each curve is averaged over 125 structures. The drug intercalated dsDNA base pairs have a higher transmission probability than that of the bare dsDNA, despite the higher rise between the two base pairs. Also, the DOS of the intercalated base pairs is higher than the corresponding bare dsDNA base pairs in the region close to the Fermi energy. These results lead to the argument that upon drug intercalation, the increase in transmission is only due to the alignment of the base pairs, since the distance between the two base pairs for the drug intercalated dsDNA is almost double than that for the bare dsDNA.      

Harashima et al.\cite{harashima2017single} reported that the conductance of a 8 bp dsDNA of sequence d-(GCTTGTTG) increases four folds upon ethidium intercalation. To have a quantitative  comparison with this experimental work\cite{harashima2017single}, we also simulated and calculated the charge transport properties of the same dsDNA sequence as used in the experiment. Fig. \ref{fig:4}E shows that the transmission increases upon intercalation with ethidium that consequently results in the higher magnitude of current as shown in Fig. \ref{fig:4}F. Our calculation shows that the magnitude of the current in ethidium-intercalated dsDNA increases about five times compared to the bare dsDNA which is in close agreement with experimental observation where four-fold increase in current upon intercalation is reported. However, the magnitude of current is lower in our calculations which is just a manifestation of different electrode couplings used. Hence, qualitatively the trend of increase in dsDNA conductance upon drug intercalation is unchanged. This signifies the robustness of our result of the enhancement in dsDNA conductance upon drug intercalation. A marked increase in the order of magnitude of dsDNA conductance is seen regardless of the dsDNA sequence studied in this work.

In conclusion, we have studied the effects of drug intercalation on the charge transport properties of a dsDNA using a multiscale modelling approach which allows to directly mimic the single-molecule conductance experimental scenarios. We find that drug intercalation increases the coherent conductance properties of dsDNA as much as by one order of magnitude. This increase is attributed to the structural changes in the dsDNA upon drug intercalation. The base pairs adjacent to the intercalator become less twisted as compared to that of bare dsDNA. This leads to the ease of charge transport through the intercalated dsDNA complexes. Therefore, any intercalation reducing the twist angle of dsDNA can increase the dsDNA conductance. The increase in conductance is found to be independent of the position of the intercalation site in the dsDNA. We also find that increasing the concentration of intercalators increases the dsDNA conductance, which provides an excellent tool to fine-tune the dsDNA conductance properties as a molecular wire. This property will be useful in developing strategies to increase the drug accumulation near DNA molecules for drug-delivery applications. Our study also provides a tool to profile the presence of intercalation in a dsDNA. We believe that an understanding of charge transport phenomenon in a drug intercalated dsDNA is paramount in studying their role in various cell functions and will eventually help to treat numerous diseases. This study advances the understanding of drug-DNA interactions that may lead to the development of anticancer, antibiotics as well as antiviral therapeutic agents in future.

%\begin{center} {\bf Acknowledgements } \end{center}
We thank Prof. Ravindra Venkatramani for his insightful remarks. A.A. and A.K.S. thank MHRD, India for the research fellowship. Authors also acknowledge DST, India, for the computational support through TUE-CMS machine, IISc. 

% Bibliography
\bibliographystyle{apsrev4-2}
\bibliography{references}

\end{document}

% The following information is for internal review, please remove them for submission
%\widetext
%\leftline{Version 04 as of \today}

% the following line is for submission, including submission to the arXiv!!
%\hspace{5.2in} \mbox{Fermilab-Pub-04/xxx-E}

\title{Fine-tuning the DNA conductance by intercalation of drug molecules: Supplementary Information}

\author{Abhishek Aggarwal}
\affiliation{Center for Condensed Matter Theory, Department of Physics, Indian Institute of Science, Bangalore-560012, India}

\author{Anil Kumar Sahoo}
\affiliation{Center for Condensed Matter Theory, Department of Physics, Indian Institute of Science, Bangalore-560012, India}

\author{Saientan Bag}
\affiliation{Center for Condensed Matter Theory, Department of Physics, Indian Institute of Science, Bangalore-560012, India}

\author{Veerabhadrarao Kaliginedi}
\affiliation{Department of Inorganic and Physical Chemistry, Indian Institute of Science, Bangalore 560012}

\author{Manish Jain}
\affiliation{Center for Condensed Matter Theory, Department of Physics, Indian Institute of Science, Bangalore-560012, India}

\author{Prabal K. Maiti}
\email{Corresponding Author: maiti@iisc.ac.in}
\affiliation{Center for Condensed Matter Theory, Department of Physics, Indian Institute of Science, Bangalore-560012, India}      
\date{\today}

\pacs{}
\maketitle

\section{MD Simulation Details}

The crystal structures of ethidium and/or daunomycin intercalated dsDNA complexes are obtained from Refs. [1] and [2], respectively. As described in our recent publication[3], we extend the DNA strands of each of the crystal structures to build 12 and 8 base pairs (bp) long dsDNAs of sequences (GCGCA$\bf{CG}$TGCGC)$_2$ and (GCA$\bf{CG}$TGC)$_2$, keeping an intercalator (daunomycin or ethidium) between the middle two base pairs (bold characters in the sequences) as shown in Fig. 2. For the above 8 bp long dsDNA, we increase the concentration of ethidium as well as  intercalate ethidium at asymmetric positions of the DNA sequence,  to build (G$\bf{CA}$CG$\bf{TG}$C)$_2$, (G$\bf{CA}$CGTGC)$_2$ and (GCACG$\bf{TG}$C)$_2$, as shown in Fig 3 A,B of the main text. Additionally, for a direct comparison with the experiment [4], we build an 8 bp long dsDNA of sequence (GCT$\bf{TG}$TTG)$_{2}$ in the presence of an intercalated ethidium molecule. We build bare dsDNA in the B-form for each of the above sequences by using the NAB tool [5]. We use the xleap module of the AMBER17 tools[6] to solvate each complex in a large rectangular box with TIP3P [7] water model. Charge neutrality of a simulation box is maintained by adding appropriate numbers of Na+ and Cl– ions, for which Joung/Cheatham ion parameters are used [8]. The AMBER ff14SB [9] with the parmbsc0 corrections [10] and GAFF [11] parameters are used for DNA and the intercalators, respectively. Further details about the molecular modeling of the intercalators can be found in Ref. [3] .

We follow, here, the same MD simulation protocol as described in our earlier publication[3]. The PMEMD module of the AMBER14 software [12] is used for performing the MD simulations. The various inter base pair parameters of the dsDNA were computed using curves+ software package[13].

\clearpage
\section{Landauer Formalism}

GAUSSIAN 09 software package[14] is used to obtain the Hamiltonian matrices for these structures using the semi-empirical method PM3[15]. The Fock matrix obtained after semi-empirical calculation, which is in the basis of atomic orbitals, is taken as the Hamiltonian matrix for subsequent calculations.
The transmission probability of the DNA molecule is calculated using the NEGF framework. The effect of the virtual electrodes attached to the dsDNA molecule is considered using the modified molecular Green’s function given by:
\begin{equation}
        G(E) = \frac{1}{(E\textbf{I}-\textbf{H}-\Sigma^{l}-\Sigma^{r})}
\end{equation}

Here,  $H$  is the Hamiltonian of the isolated molecule. The self-energies  $\Sigma^{l}$ and $\Sigma^{r}$ describe the effect of the left (l) and right (r) electrodes respectively, on the broadening in the molecular energies. The transmission probability for charge transport from one electrode to the other electrode over all the pathways is given by:

\begin{equation}
        T(E) = \Gamma^{l}G\Gamma^{r}G^{\dagger}
\end{equation}

Here, $\Gamma^{l}$ and $\Gamma^{r}$ are the broadening matrices given by $\Gamma = i[\Sigma - \Sigma^{\dagger}]$. Only the imaginary part of the broadening matrix is considered in our calculations as has also been used in several charge transport works[16-18]. The electrode atoms are not explicitly modelled; instead, the broadening matrices are used to consider the effect of the electrodes. We assume that the electrodes and the linkers affect only the terminal base pairs and add the broadening parameter on the orbitals representing the terminal base pairs’ atoms only. Hence, the elements of the broadening matrices are given as $\Gamma_{ij}  = 0.1$ eV, for the terminal base pair atomic orbitals and $i = j$, and is taken as 0 eV otherwise.

Using the above formalism and parameters, we get the value of the transmission coefficient for a range of energy values. The effect of using different broadening parameter is explored in next section. Landauer expression is used to get the value of current I at a given applied potential V:
\begin{equation}
        I = \frac{2e^{2}}{h}\int_{\infty}^{-\infty} dE[f(E + \frac{eV}{2}) - f( E - \frac{eV}{2})]T(E)
\end{equation}

Here, $f(E)$  is the Fermi energy function and is given by:
\begin{equation}
        f(E) = \frac{1}{1 + exp((E-\mu)/k_{B}T)}
\end{equation}

where $k_{B}$  is the Boltzmann constant, $T$  is the temperature taken as 300 K, and $\mu$ is the chemical potential of the electrodes. Bag et al. find that the Fermi level of the dsDNA system increases by 0.36 eV upon the attachment of gold electrode[19, 20]. Hence, $\mu$ is taken as 0.36 eV above the HOMO energy level of the dsDNA/drug-dsDNA complex system.

\clearpage

\section{Discussion on validity of Landauer formalism}

We study the charge transport properties of small dsDNA sequences of 8 base pairs length. As shown in various previous studies[21-23], tunneling phenomenon is the dominant charge transport process in short dsDNA sequences, which justifies to use tunneling charge transport mechanism under the framework of Landauer theory. 

In our very recent work[20], we have discussed the time scales involved in the base pair dynamics of dsDNA and the charge propagation in dsDNA. We found that the base pair structural parameters are correlated on a time scale of nanoseconds, while the charge transport is correlated on sub-picosecond time scale. Several other works[24, 25] have explored the time scales involved in the dynamics and charge propagation and found the time scales of the same order. The dsDNA structures chosen within a time interval of picoseconds will be correlated to each other and there will be a dynamic disorder, where the electron-phonon interactions will become important. However, in this work, the structures chosen for the charge transfer study are at least 40 ps apart which ensure that the structures are correlated dynamically but are static with respect to charge transfer. This backs the assumption that structural changes in DNA as static disordered images, over which an average can be carried out.

\clearpage

\section{Effect of Coupling Parameter}

The electronic coupling values between the electrode and the molecule depends on a variety of factors[26] and can vary depending on the arrangement and material of the electrodes and the linkers. Generally, the magnitude of the electrode couplings is found to be of the order of 0.1 – 3.0 eV depending on the linkers and the material of electrodes[26, 27]. We have now performed rigorous calculations to motivate the use of coupling parameter values. We have used different numerical values of electrode couplings for the calculations ranging from 0.1 eV to 5.0 eV as shown in Fig. S1. For all the used values for this parameter, we see that the results only change quantitively and not qualitatively.

\begin{figure}[htbp]
    \centering
        \includegraphics[width=\columnwidth]{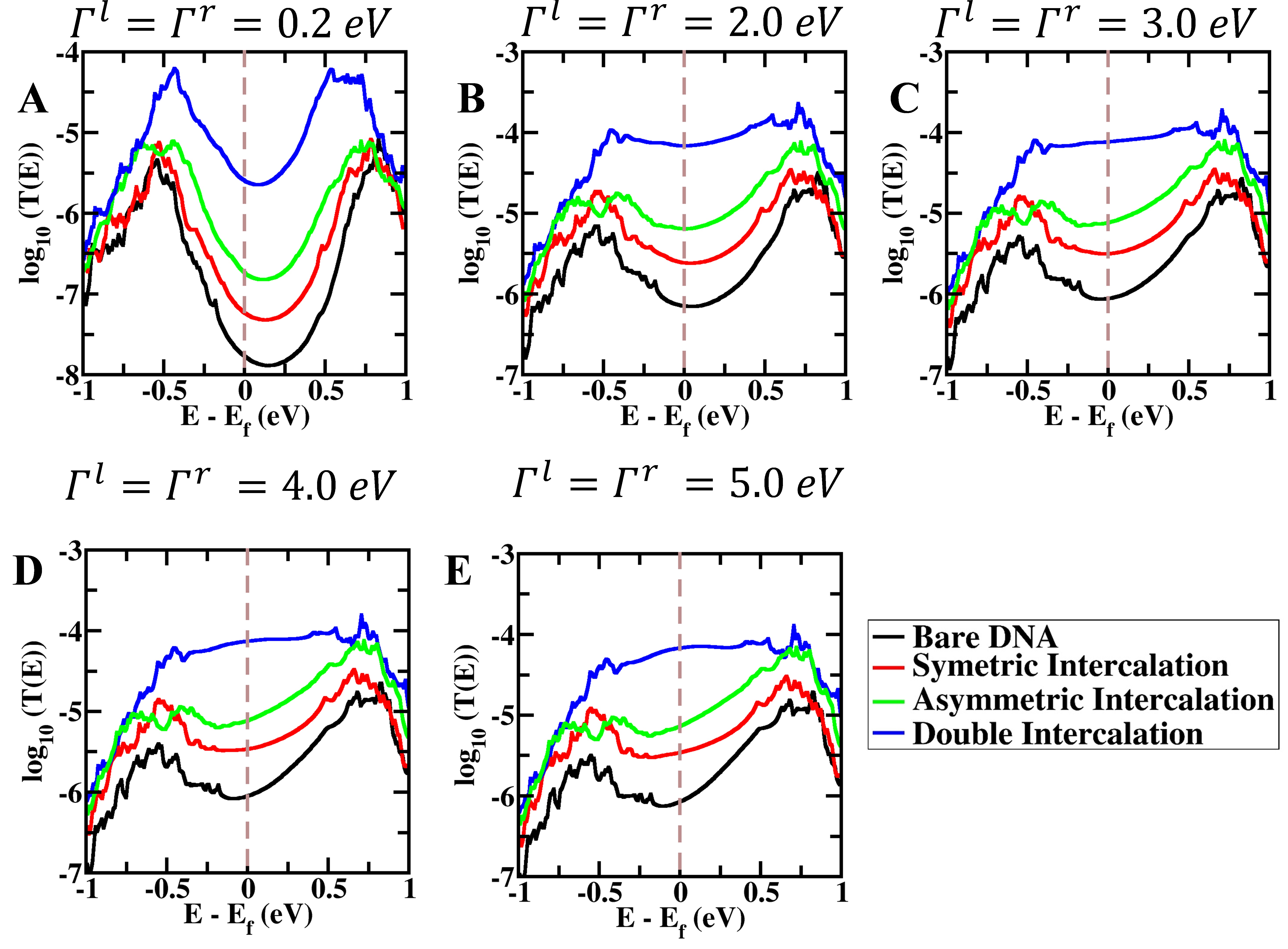}
	\caption{Transmission probability curve for the DNA and drug-DNA complexes (averaged over 50 morphologies) in the region close to the Fermi-energy for 8 bp dsDNA sequence with and without intercalators for various electrode coupling values.}

\end{figure}

In the real-world STM-BJ experimental setups, the coupling at the two ends can sometimes be asymmetric because of the different attachments of the linkers to the electrodes. To investigate this, we have also computed the transmission through intercalated as well as bare dsDNA with asymmetric couplings at applied at the two ends as shown in Fig. S2. We fixed the electrode coupling value for left electrode at 0.1 eV, while varied the value for another electrode from 0.0001 eV to 10 eV (Fig. R2). We chose these extreme limits of electrode coupling values to incorporate all possible asymmetries in the setup and find that for each case, the relative trend of the transmission of intercalated vs non-intercalated dsDNA molecules remains same and only the magnitude of the transmission changes. The conclusion of the work remains same, i.e. DNA conductance enhances upon intercalation. 

\begin{figure}[htbp]
    \centering
        \includegraphics[width=\columnwidth]{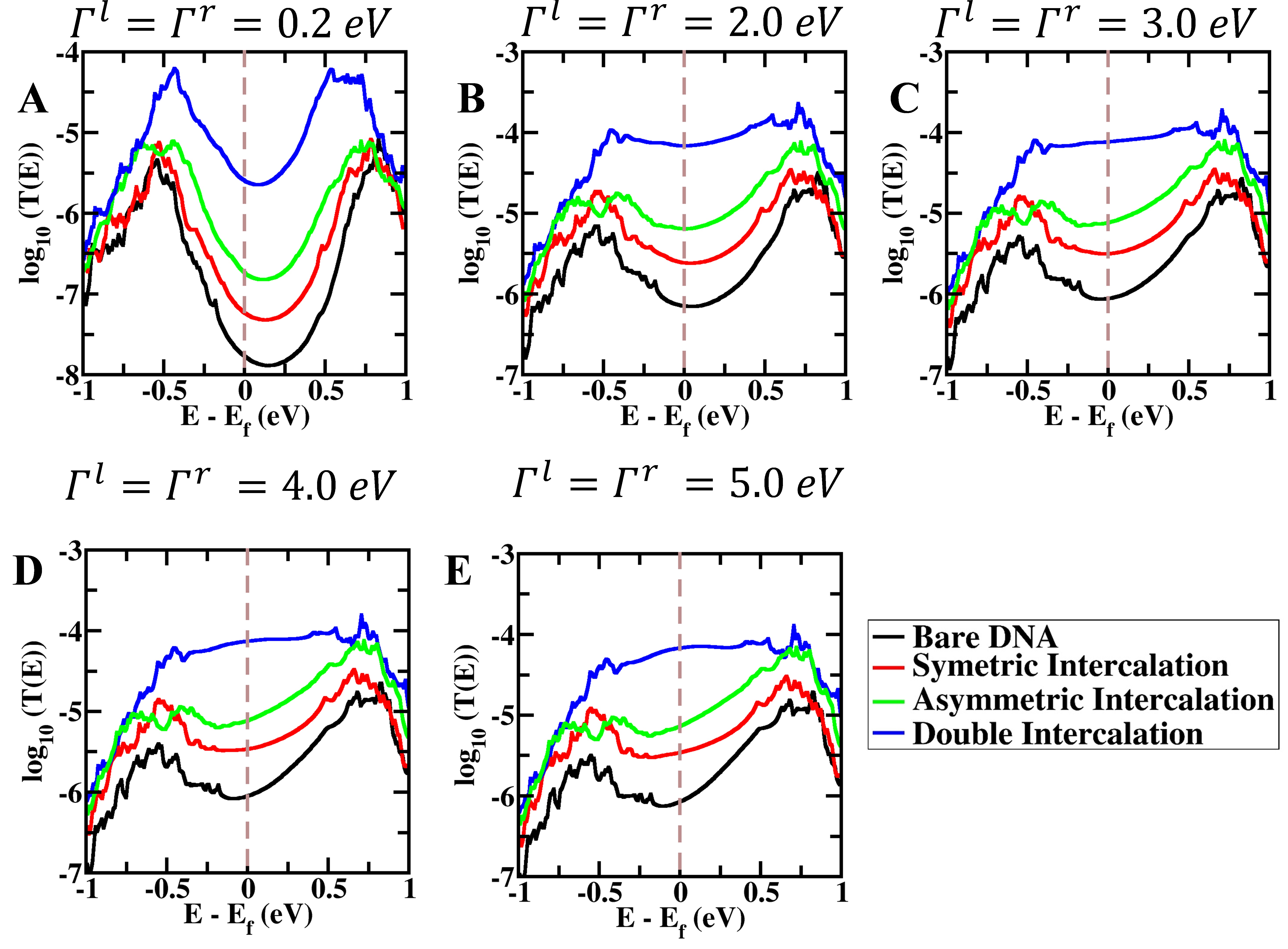}
        \caption{Transmission probability curve for the DNA and drug-DNA complexes (averaged over 50 morphologies) in the region close to the Fermi-energy for 8 bp dsDNA sequence with and without intercalators for various asymmetric electrode coupling values. The left electrode coupling is kept fixed at $\Gamma_{l}$ = 0.1 eV, while $\Gamma_{r}$ is varied from 0.0001 eV to 10 eV. In all the cases, intercalated dsDNA has higher transmission than bare dsDNA.}
   
\end{figure}

\clearpage

\section{Effect of interaction between electrodes and molecule}

The attachment of the metal leads to the molecule may adjust the position of the frontier molecular orbitals which can consequently change the Fermi level of the system. In this work, in all the calculations, the intercalator is 2 or 4 base pairs away from the electrodes which is around 7 $\AA$ and 14 $\AA$ from the electrodes. This distance is very high to affect the charge transfer between the molecule and the electrodes. So, the readjustment of the frontier molecular orbitals can be considered similar for both the intercalated as well as bare dsDNA. This argument is further justified when different electrode couplings are used for ethidium intercalated asymmetrically into the dsDNA (as shown in Fig. 2 of the main text). Here, we apply different electrode couplings to the terminals of bare dsDNA as well as dsDNA intercalated with ethidium to check whether the observed trends are robust to the electrode coupling values. As shown in Fig. S2, in each case, after intercalation the dsDNA conductance increases regardless of the electrode couplings used. Notice the increase of magnitude of transmission as the coupling values are increased.

To check the effect of the choice of different Fermi energies, we calculate the current of the bare dsDNA as well as intercalated dsDNA for 75 morphologies each at an applied potential bias of 1 V with the Fermi Energy taken as a range of energies near the HOMO level of the molecules (Fig. S3) and show that for all the Fermi energies, the intercalated dsDNA has a higher average conductance than bare dsDNA.

\begin{figure}[htbp]
    \centering
        \includegraphics[width=0.5\columnwidth]{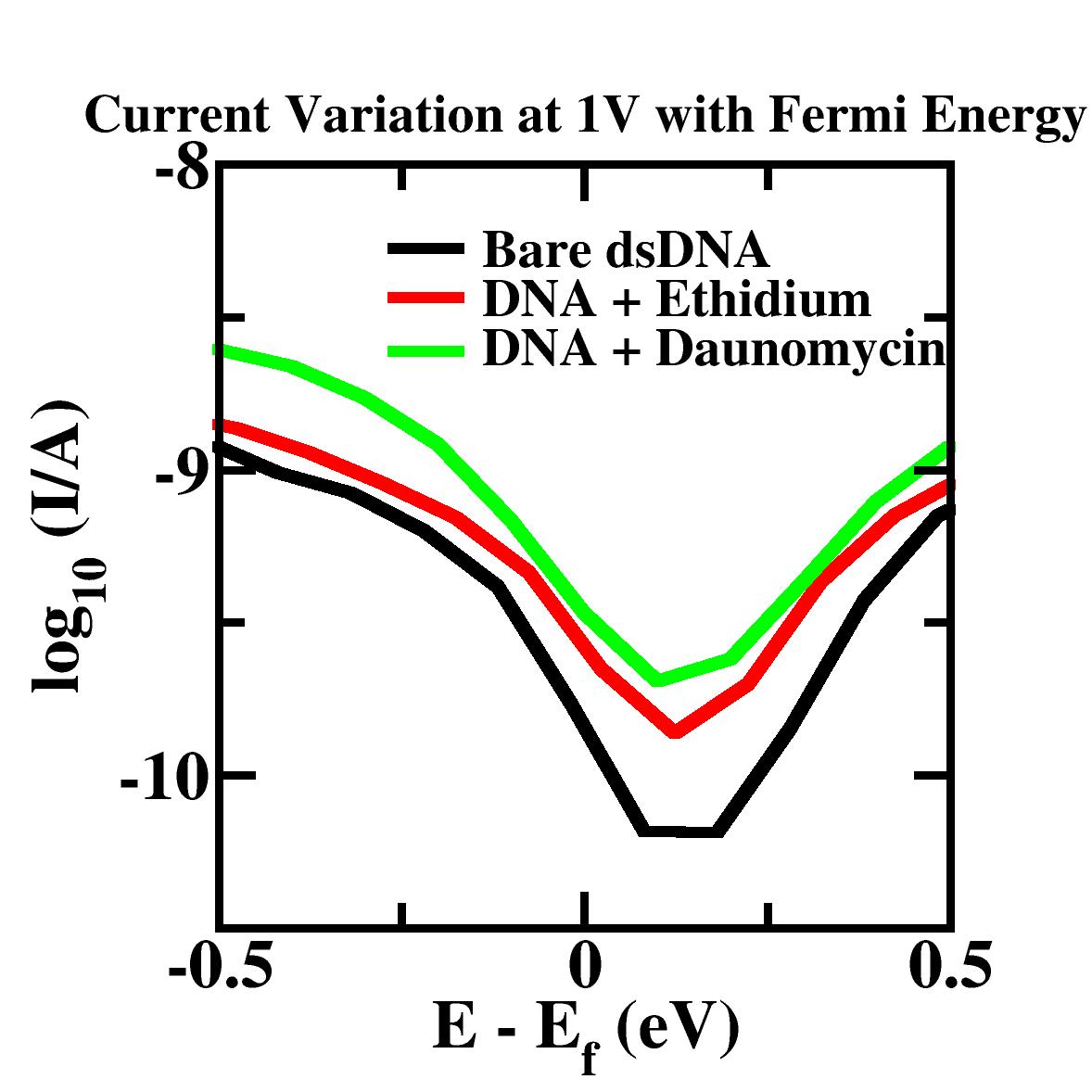}
        \caption{Variation of current at 1 V vs. Fermi energy averaged over 75 morphologies of bare dsDNA and dsDNA intercalated with ethidium and daunomycin. At any particular Fermi Energy, intercalated dsDNA has higher current at 1 V relative to bare dsDNA.}
    
\end{figure}

\clearpage

\section{Structural Parameters of 8 bp and 12 bp intercalated dsDNA}

We calculate different inter-base pair helical parameters of the 12 bp bare dsDNA as well as intercalated dsDNA complexes in Fig. S1. For each of the intercalated ligand–DNA complexes, the rise of the base pair step at the intercalation site is almost double of that of the bare dsDNA; otherwise, the rise is the similar for rest of the base pairs irrespective of the presence of an intercalator similar to the 8 bp dsDNA and intercalated dsDNA molecules. The slide also varies in the intercalated region. The magnitude of twist angle for the intercalated region is significantly less than that of the bare dsDNA. This signifies the unwinding of the base pairs in the intercalated region upon intercalation in the 12 bp dsDNA as well upon intercalation. Thus, the base pairs adjacent to the intercalators have high rise but at the same time a lower twist angle than the corresponding normal dsDNA base pairs.

\begin{figure}[h]
    \centering
	\includegraphics[width=\columnwidth]{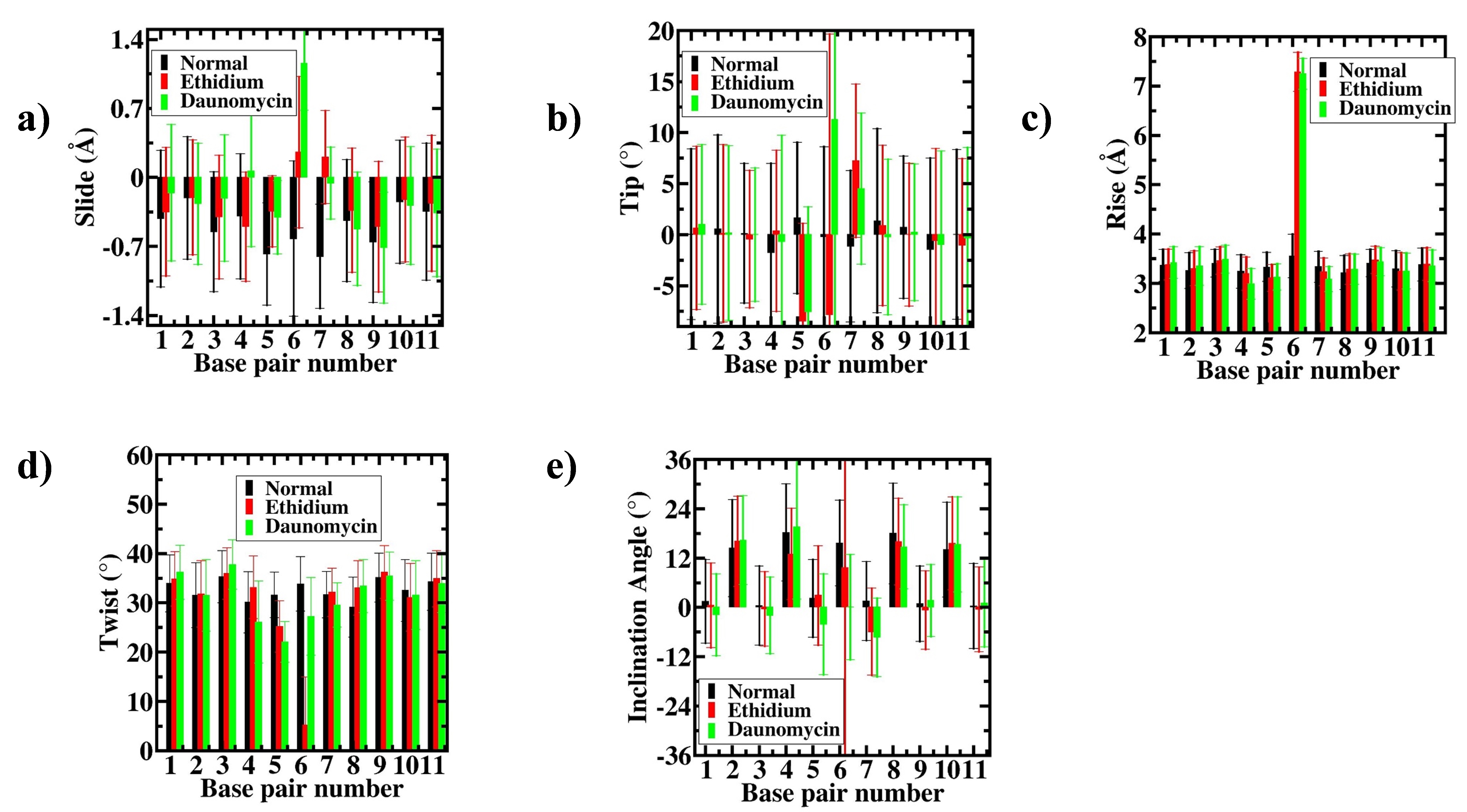}
    \caption{A comparison of structural parameters of 12 bp dsDNA with and without intercalator.}
    
\end{figure}

For 8 bp dsDNA, apart from structural parameters shown in main text, fig. S5 shows the other structural parameters.

\begin{figure}[htbp]
    \centering
        \includegraphics[width=0.75\columnwidth]{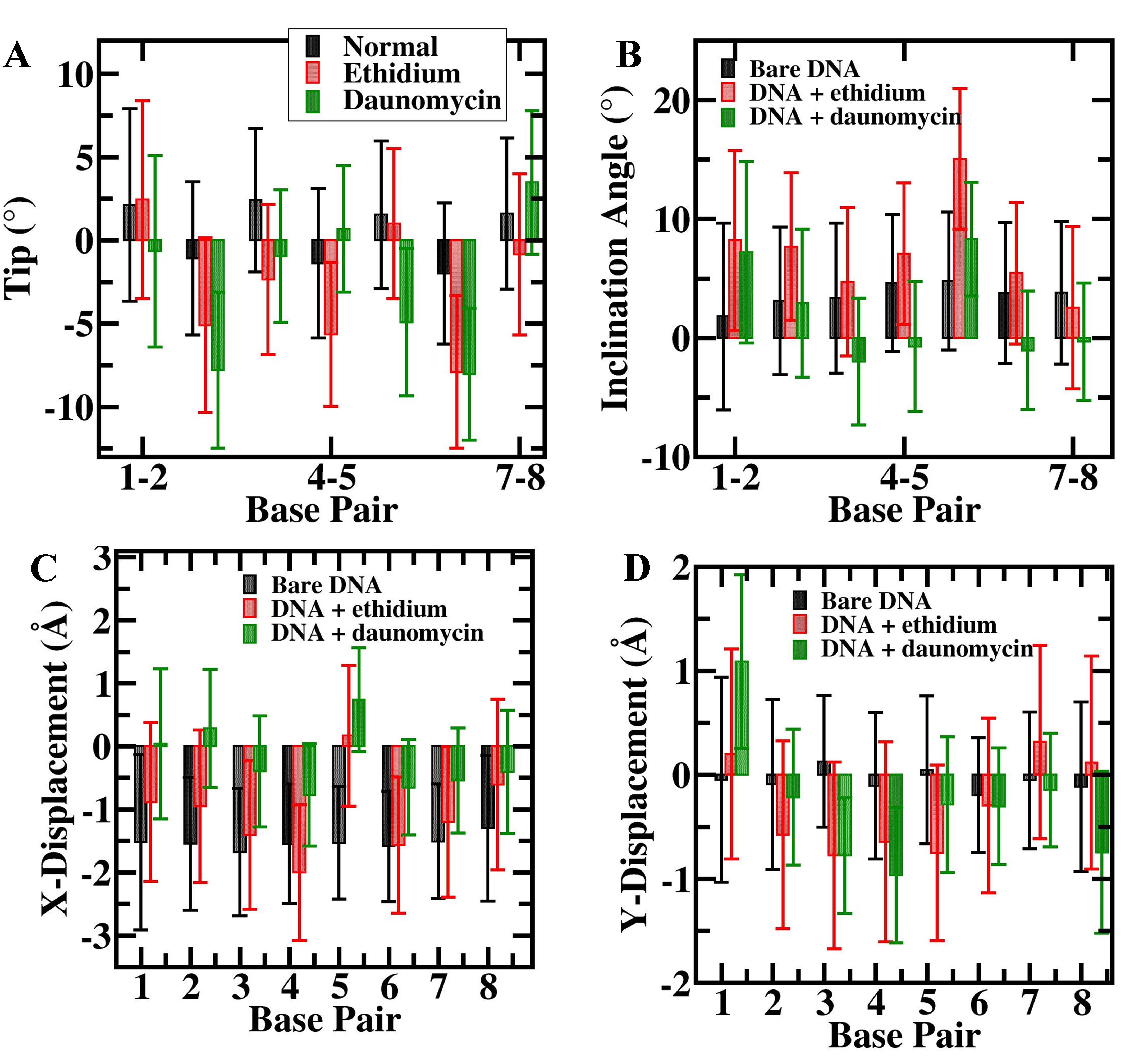}
    \caption{A comparison of structural parameters of 8 bp dsDNA with and without intercalator.}
   
\end{figure}

\clearpage

\section{Charge transport properties of 12 bp intercalated dsDNA}

The charge transport properties of 12 bp dsDNA change drastically upon small-molecule intercalation. Fig. S2 (a) and (b) show the comparison of I-V characteristic and transmission probabilities for the 12 bp dsDNA intercalated with ethidium or daunomycin and the normal 12 bp dsDNA for a range of energies near the Fermi energy of the molecule. Clearly, the intercalated dsDNA has higher transmission and hence higher conductance relative to the normal dsDNA for both the intercalating ligands. The daunomycin–dsDNA complex has almost one order of magnitude higher conductance than the normal dsDNA. 

\begin{figure}[htbp]
    \centering
        \includegraphics[width=\columnwidth]{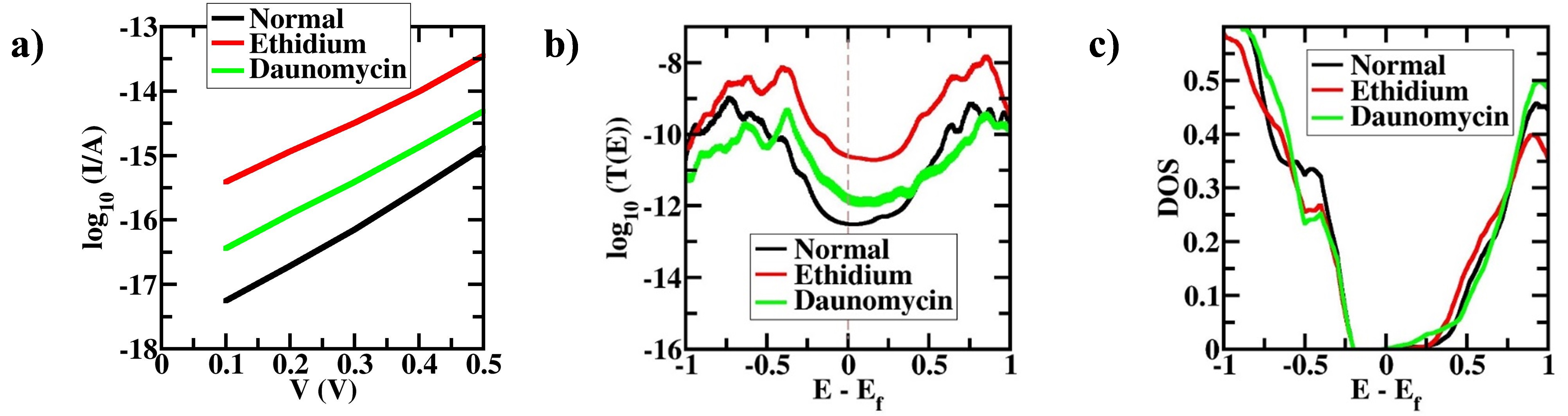}
    \caption{a) V-I characteristics curves of 12 bp dsDNA with and without intercalators. Clearly, the current increases by orders of magnitude upon intercalation. b) Transmission probability curve in the near Fermi-energy region for 12 bp dsDNA sequence with and without intercalators. c) Density of States for 12 bp dsDNA. Clearly, intercalated dsDNA has a lower HOMO-LUMO gap relative to the normal dsDNA. Due to this, the transmission probabilities are also higher for intercalated dsDNA.}
   
\end{figure}

\clearpage

\section{Role of Backbone in DNA Charge Transfer}
There has been a recent discussion about the possibility of having electronic transport mediated by the backbones[28] which motivated us to investigate the role of backbone in present study. In Fig. S7 a)-c) and d)-e), we present the isosurface of HOMO orbitals of three randomly chosen structures of a bare dsDNA, as well as ethidium and daunomycin intercalated dsDNA for 8 bp and 12 bp dsDNA sequences respectively. Clearly, the HOMO isosurfaces are localized over both the bases as well as backbone and thus the backbone contributes to the charge transport in dsDNA and intercalated dsDNA. 

\begin{figure}[htbp]
    \centering
        \includegraphics[width=\columnwidth]{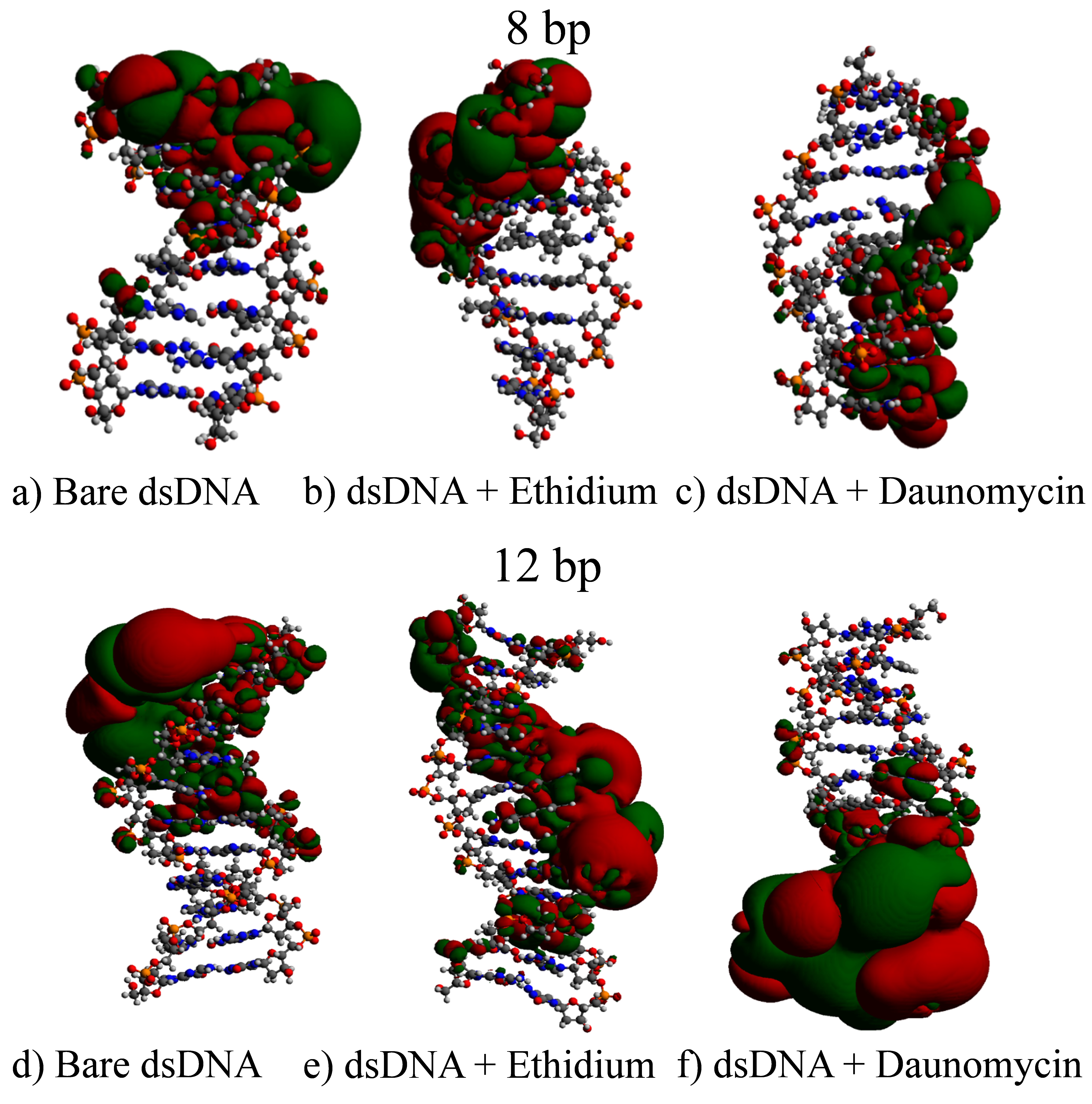}
	\caption{HOMO level distribution on 8 bp a) bare dsDNA b) dsDNA intercalated with ethidium, c) dsDNA intercalated with daunomycin, and 12 bp e) bare dsDNA f) dsDNA intercalated with ethidium, g) dsDNA intercalated with daunomycin.}

\end{figure}

To check the effect of backbone on the relative trend of the conductance of dsDNA change upon intercalation, we calculate the charge transport properties using the same 50 structures without using backbone as shown in Fig. S8. Clearly, without backbone, the dsDNA conductance decreases upon intercalation indicating the clear role of backbone in the charge transport of dsDNA.

\begin{figure}[htbp]
    \centering
        \includegraphics[width=0.5\columnwidth]{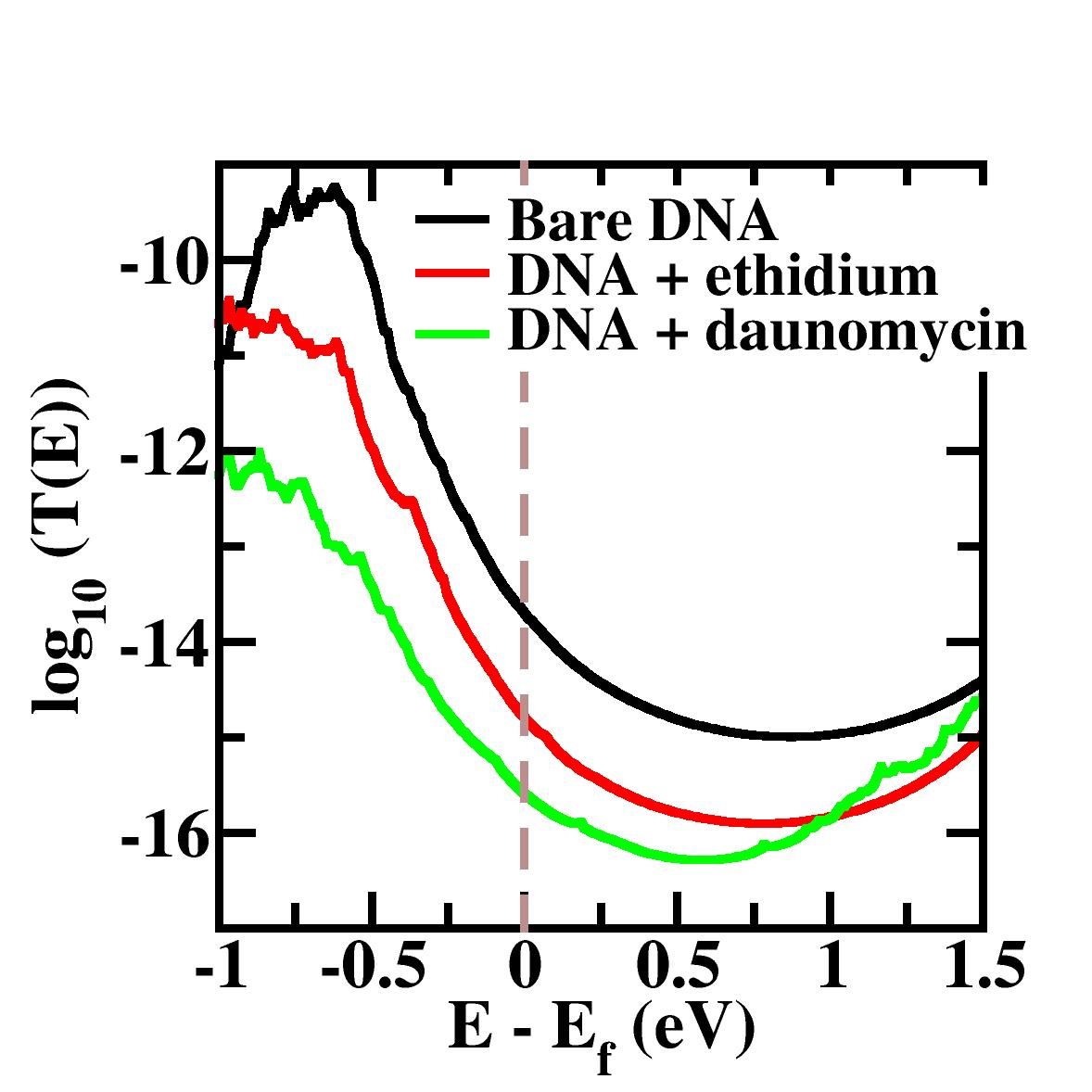}
	\caption{Transmission probability curve for the DNA and drug-DNA complexes (averaged over 50 morphologies) in the region close to the Fermi-energy for 8 bp dsDNA sequence with and without intercalators without backbone included in the calculations. This plot signifies the importance of backbone in charge transport in dsDNA.}

\end{figure}

\clearpage

\section*{References}

1.	Jain, S.C. and H.M. Sobell, Visualization of drug-nucleic acid interactions at atomic resolution. VIII. Structures of two ethidium/dinucleoside monophosphate crystalline complexes containing ethidium: cytidylyl (3'-5') guanosine. Journal of Biomolecular Structure and Dynamics, 1984. 1(5): p. 1179-1194.

2.	Wang, A.H., et al., Interactions between an anthracycline antibiotic and DNA: molecular structure of daunomycin complexed to d (CpGpTpApCpG) at 1.2-. ANG. resolution. Biochemistry, 1987. 26(4): p. 1152-1163.

3.	Sahoo, A.K., B. Bagchi, and P.K. Maiti, Understanding enhanced mechanical stability of DNA in the presence of intercalated anticancer drug: Implications for DNA associated processes. The Journal of chemical physics, 2019. 151(16): p. 164902.

4.	Harashima, T., et al., Single-molecule conductance of DNA gated and ungated by DNA-binding molecules. Chemical Communications, 2017. 53(75): p. 10378-10381.

5.	Leontis, N.B., J. SantaLucia, and A.C.S. Staff, Molecular modeling of nucleic acids. 1998: ACS Publications.

6.	Case, D., et al., Duke, TJ Giese, H. Gohlke, AW Goetz, N. Homeyer, S. Izadi, P. Janowski, J. Kaus, A. Kovalenko, TS Lee, S. LeGrand, P. Li, T. Luchko, R. Luo, B. Madej, KM Merz, G. Monard, P. Needham, H. Nguyen, HT Nguyen, I. Omelyan, A. Onufriev, DR Roe, A. Roitberg, R. Salomon-Ferrer, CL Simmerling, W. Smith, J. Swails, RC Walker, J. Wang, RM Wolf, X. Wu, DM York and PA Kollman Amber, 2015.

7.	Jorgensen, W.L., et al., Comparison of simple potential functions for simulating liquid water. The Journal of chemical physics, 1983. 79(2): p. 926-935.

8.	Joung, I.S. and T.E. Cheatham III, Determination of alkali and halide monovalent ion parameters for use in explicitly solvated biomolecular simulations. The journal of physical chemistry B, 2008. 112(30): p. 9020-9041.

9.	Maier, J.A., et al., ff14SB: improving the accuracy of protein side chain and backbone parameters from ff99SB. Journal of chemical theory and computation, 2015. 11(8): p. 3696-3713.

10.	Perez, A., et al., Refinement of the AMBER force field for nucleic acids: improving the description of $\alpha$/$\gamma$ conformers. Biophysical journal, 2007. 92(11): p. 3817-3829.

11.	Wang, J., et al., Development and testing of a general amber force field. Journal of computational chemistry, 2004. 25(9): p. 1157-1174.

12.	Case, D.A., et al., The Amber biomolecular simulation programs. Journal of computational chemistry, 2005. 26(16): p. 1668-1688.

13.	Lavery, R., et al., Conformational analysis of nucleic acids revisited: Curves+. Nucleic acids research, 2009. 37(17): p. 5917-5929.

14.	Frisch, M., et al., Gaussian 09, revision a. 02, gaussian. Inc., Wallingford, CT, 2009. 200: p. 28.

15.	Stewart, J.J., Optimization of parameters for semiempirical methods II. Applications. Journal of computational chemistry, 1989. 10(2): p. 221-264.

16.	Seth, C., et al., Conductance in a bis-terpyridine based single molecular breadboard circuit. Chemical science, 2017. 8(2): p. 1576-1591.

17.	Gutiérrez, R., et al., Inelastic quantum transport in a ladder model: Implications for DNA conduction and comparison to experiments on suspended DNA oligomers. Physical Review B, 2006. 74(23): p. 235105.

18.	Qi, J., et al., Unified model for conductance through DNA with the Landauer-Büttiker formalism. Physical Review B, 2013. 87(8): p. 085404.

19.	Bag, S., et al., Anisotropic Charge Transport in Nanoscale DNA Wire. The Journal of Physical Chemistry C, 2020.

20.	Aggarwal, A., et al., Multiscale modelling reveals higher charge transport efficiencies of DNA relative to RNA independent of mechanism. Nanoscale, 2020.

21.	Li, Y., et al., Thermoelectric effect and its dependence on molecular length and sequence in single DNA molecules. Nature communications, 2016. 7(1): p. 11294.

22.	Genereux, J.C. and J.K. Barton, Mechanisms for DNA charge transport. Chemical reviews, 2009. 110(3): p. 1642-1662.

23.	Guo, C., et al., Molecular rectifier composed of DNA with high rectification ratio enabled by intercalation. Nature chemistry, 2016. 8(5): p. 484.

24.	Troisi, A., A. Nitzan, and M.A. Ratner, A rate constant expression for charge transfer through fluctuating bridges. The Journal of chemical physics, 2003. 119(12): p. 5782-5788.

25.	Troisi, A. and G. Orlandi, Hole migration in DNA: a theoretical analysis of the role of structural fluctuations. The journal of physical chemistry B, 2002. 106(8): p. 2093-2101.

26.	Jia, C. and X. Guo, Molecule–electrode interfaces in molecular electronic devices. Chemical Society Reviews, 2013. 42(13): p. 5642-5660.

27.	Chen, F., et al., Effect of anchoring groups on single-molecule conductance: comparative study of thiol-, amine-, and carboxylic-acid-terminated molecules. Journal of the American Chemical Society, 2006. 128(49): p. 15874-15881.

28.	Zhuravel, R., et al., Backbone charge transport in double-stranded DNA. Nature nanotechnology, 2020: p. 1-5.

\bibliographystyle{apsrev4-2}